\documentclass[letterpaper]{natureprintstyle}

\usepackage{graphicx}
\usepackage{amsmath, amssymb}
\usepackage{multirow}
\usepackage[varg]{txfonts}

\newcommand{\capbf}[1]{\textbf{\textsf{#1}}}

\title{Delocalization of a disordered bosonic system by repulsive interactions}

\author{B. Deissler$^{1}$, M. Zaccanti$^{1}$, G. Roati$^{1}$, C. D'Errico$^{1}$, M. Fattori$^{1,2}$, M. Modugno$^{1}$, G. Modugno$^{1}$ \& M. Inguscio$^{1}$}

\begin{document}

\maketitle

\begin{affiliations}
 \item LENS and Dipartimento di Fisica, Universit\`a di Firenze, and CNR-INFM, 50019 Sesto Fiorentino, Italy
 \item Museo Storico della Fisica e Centro Studi e Ricerche `E. Fermi', 00184 Roma, Italy
\end{affiliations}

\begin{abstract}
Clarifying the interplay of interactions and disorder is fundamental to the understanding of many quantum systems, including superfluid helium in porous media\cite{Reppy 92}, granular and thin-film superconductors\cite{Dubi07,Beloborodov07,Goldman98,Phillips03}, and light propagating in disordered media\cite{Pertsch04,Schwartz07,Lahini08}.
One central aspect for bosonic systems\cite{Giamarchi88,Fisher88,Scalettar91} is the competition between disorder, which tends to localize particles, and weak repulsive interactions, which instead have a delocalizing effect. Since the required degree of independent control of the disorder and of the interactions is not easily achievable in most available physical systems, a systematic experimental investigation of this competition has so far not been possible. Here we employ a degenerate Bose gas with tunable repulsive interactions in a quasi-periodic lattice potential to study this interplay in detail. We characterize the entire delocalization crossover through the study of the average local shape of the wavefunction, the spatial correlations, and the phase coherence. Three different regimes are identified and compared with theoretical expectations\cite{Lugan07,Roux08,Lee90,Damski03,Schulte05,Falco09}: an exponentially localized Anderson glass, the formation of locally coherent fragments, as well as a coherent, extended state. Our results illuminate the role of weak repulsive interactions on disordered bosonic systems and show that the system and the techniques we employ are promising for further investigations of disordered systems with interactions, also in the strongly correlated regime\cite{Fisher89,Fallani07,White09,Bloch08}.

\end{abstract}

The interplay of disorder and interactions lays at the heart of the behaviour of many physical systems. Notable examples are the transitions to insulators observed in superconductors and metals\cite{Dubi07,Beloborodov07,Goldman98,Phillips03}, quantum Hall physics\cite{Martin09}, electrical conduction in DNA\cite{Endres04}, or light propagation in non-linear disordered media\cite{Schwartz07,Lahini08}. An important step towards their full comprehension is understanding disordered bosonic systems at zero temperature, where a competition between disorder and weak repulsive interactions is expected. Indeed, while disorder tends to localize non-interacting particles giving rise to Anderson localization\cite{Anderson58}, weak repulsive interactions can counteract this localization in order to minimize the energy. Eventually, interactions can screen the disorder and bring the system towards a coherent, extended ground state, i.e.\ a Bose-Einstein condensate (BEC). In many years of research, mainly theoretical predictions have been made about the properties of the complex phases expected to appear as a result of this competition\cite{Giamarchi88,Scalettar91,Fisher88,Lugan07,Roux08,Lee90,Damski03,Schulte05,Falco09}.
A systematic experimental study has so far not been possible, since on the one hand interactions in condensed matter systems are strong but difficult to control\cite{Reppy92}, while on the other hand in photonic systems only non-linearities corresponding to attractive interactions\cite{Schwartz07,Lahini08} have been explored in experiments. Instead, ultracold atoms in disordered optical potentials are a promising system for such investigations\cite{Damski03,Fallani08}, and have already enabled the observation of Anderson localization for bosons in the regime of negligible interactions\cite{Billy08,Roati08}. Using one of these systems in a disordered lattice, we characterize the whole crossover from the regime of disorder-induced localization to that of Bose-Einstein condensation by tuning repulsive interactions in a controlled manner. The simultaneous measurement of localization properties, spatial correlations and phase coherence properties, and the comparison with the predictions of a theoretical model allow us to identify the different regimes of this delocalization crossover.

\begin{figure}[!t]
 \begin{center}
 \vspace{6pt}
\includegraphics[width = 0.4\textwidth]{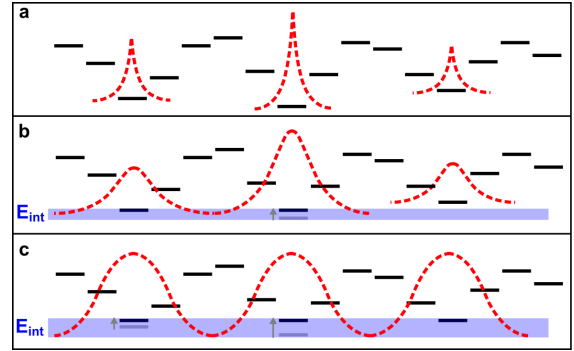}
 \end{center}
\caption{\capbf{Delocalization in a quasi-periodic potential.} Cartoon of the interaction-induced delocalization. \capbf{a},  In a very weakly interacting system with sufficiently large disorder, the eigenstates are exponentially localized, and several of the lowest energy states are populated an average of 4.4 lattice sites apart (Anderson glass). \capbf{b}, The energies of different states can become degenerate due to repulsive interactions and their shape might be modified, giving rise to the formation of locally coherent fragments (fragmented BEC), though global phase coherence is not restored until \capbf{c}, the entire system forms a coherent, extended state (BEC) at large interaction strengths.} 
\label{fig:bichromatic}
\end{figure}

The system employed consists of a three-dimensional degenerate Bose gas of ${}^{39}$K in a one-dimensional quasi-periodic potential, which is generated by perturbing a strong primary optical lattice of periodicity $d = \pi/k_1$ with a weak secondary lattice of incommensurate periodicity $\pi/k_2$ ($k=2\pi/\lambda$, where $\lambda$ is the wavelength of the light generating the lattice). The corresponding Hamiltonian is characterized by the site-to-site tunnelling energy $J$ of the primary lattice, which is kept fixed in the experiment, and the disorder strength $\Delta$. The interatomic interactions can be controlled by changing the atomic $s$-wave scattering length $a$ by means of a Feshbach resonance\cite{Roati07}, which in turn determines the mean interaction energy per particle $E_{int}$ (see Methods).

In the case of non-interacting atoms, such a system is a realization of the Aubry-Andr\'e model\cite{Aubry80}, which shows an Anderson-like localization transition for a finite value of the disorder $\Delta/J = 2$. Above the transition, the non-interacting eigenstates of the potential are exponentially localized due to the quasi-periodic perturbation of the lattice on-site energies and the energy spectrum is split into ``minibands''\cite{Roux08,Modugno09}.
The localization properties in this case have been studied experimentally in detail in ref.~\citen{Roati08}, where it was seen that several low-lying eigenstates, separated on average by $d/(\beta-1) \approx 4.4d$, where $\beta = k_2/k_1$, are typically populated in the experiment. Adding weak interactions, the different regimes that appear as a result of the interplay of disorder and interactions can be explored. For very weak repulsive interactions, the occupation of several eigenstates in the lowest miniband is favoured (Fig.~1a). This regime, in which several exponentially localized states coexist without phase coherence, is often identified with an Anderson glass\cite{Scalettar91,Damski03} (AG). 
As $E_{int}$ is increased, coherent fragments, that extend over more than one well of the quasi-periodic potential, are expected to form (Fig.~1b). In this case, global phase coherence would not yet be restored, and the local shape of the states might be modified. Some authors have called this regime a `fragmented BEC'\cite{Lugan07} (fBEC). Finally, for large enough $E_{int}$ a single, extended phase-coherent state is expected to be formed, i.e.\ a macroscopic BEC (Fig.~1c). 

\begin{figure}
\begin{center}
\includegraphics[width = 0.4\textwidth]{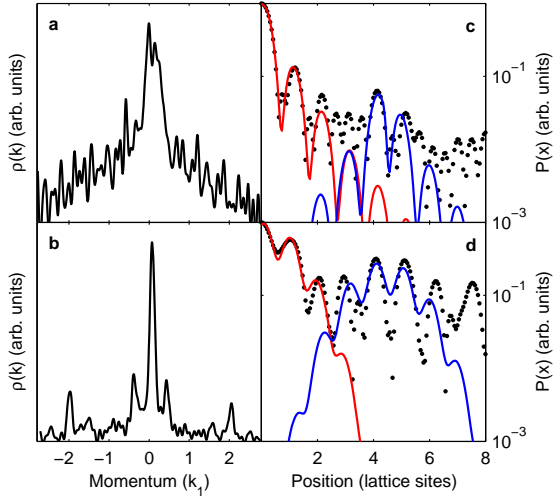}
\end{center}
\caption{\capbf{Analyzing the momentum distribution.} \capbf{a}, Typical momentum distribution and \capbf{c}, mean local shape of the wavefunction recovered from a Fourier transform (FT) in the localized regime, and \capbf{b,d} in the extended regime. The root-mean-squared width of the momentum distributions and the exponent extracted from a fit (red and blue lines) to the FT give the localization properties. The coherence properties are extracted by measuring the fluctuations of the phase of the interference pattern in the momentum distribution, or by the relative height of the two states $4.4d$ apart, which can be related to the spatially averaged correlation function $g(4.4d)$.}
\label{fig:MDcomparison}
\end{figure}

The system is prepared by first loading an interacting condensate adiabatically from the ground state of a harmonic trap into the quasi-periodic lattice. The interaction energy is then slowly changed to its final value $E_{int}$, while the confining potential is reduced. This process is adiabatic for most of the parameter range explored until $E_{int}$ becomes sufficiently low for the system to enter the fully localized regime. Here, several independent low-lying excited states are populated even when it would be energetically favourable to populate just the ground state. This loss of adiabaticity is seen experimentally as a transfer of energy into the radial direction (see Supplementary Information).

\begin{figure}[t]
\begin{center}
\includegraphics[width = 0.33\textwidth]{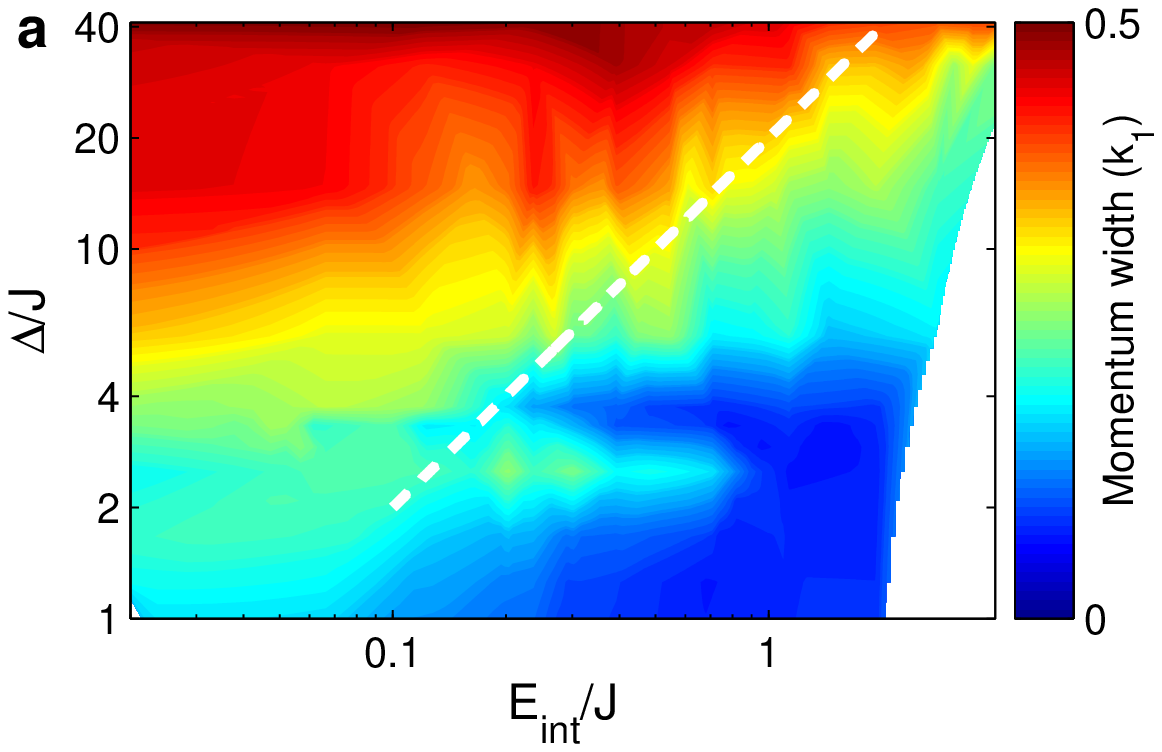}
\includegraphics[width = 0.33\textwidth]{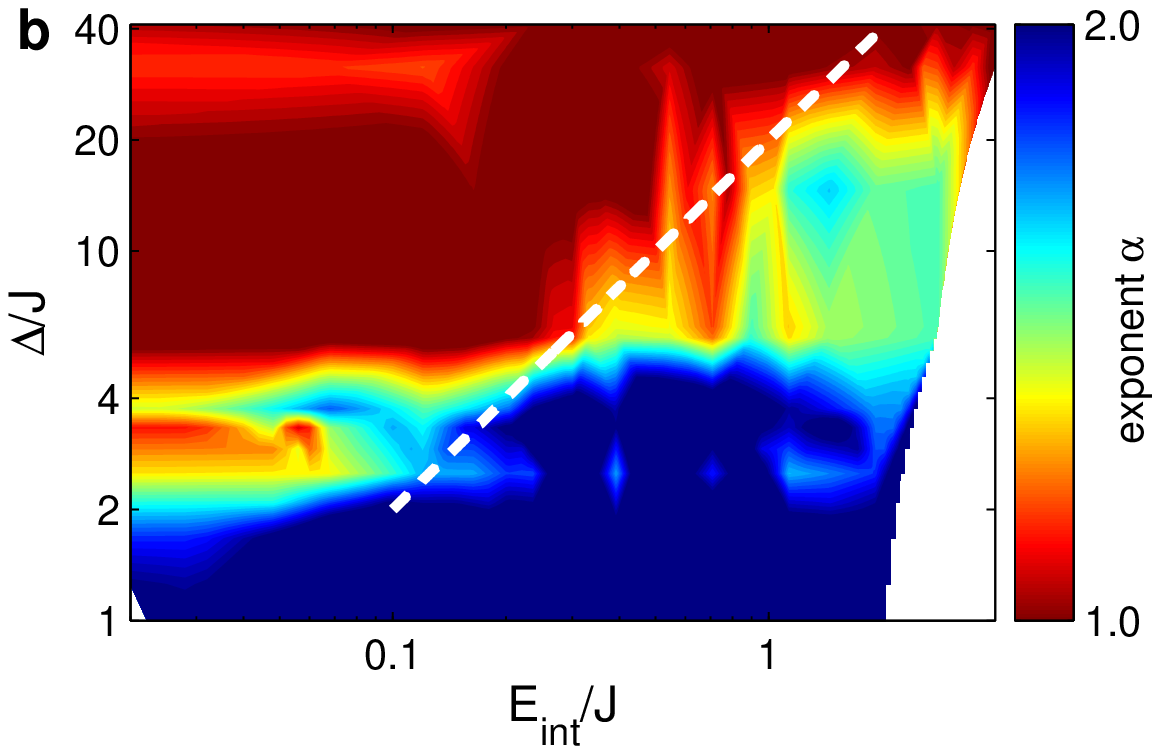}
\end{center}
\caption{\capbf{Probing the interaction-induced delocalization.} \capbf{a}, Root-mean-square width (in units of $k_1$) of central peak of momentum distribution. The white line gives $E_{int} = 0.05\Delta$ (standard deviation of the lowest-lying energies), where we expect the centre of the delocalization crossover. 
\capbf{b}, Average exponent $\alpha$ of states occupying the potential wells. The line is the same as in \capbf{a}. The data taken at different values of $\Delta/J$ and $E_{int}/J$ are linearly interpolated; the color indicates the mean value of the measured quantity.}
\label{fig:PD}
\end{figure}

\begin{figure*}[!t]
\begin{center}
\includegraphics[width = 0.33\textwidth]{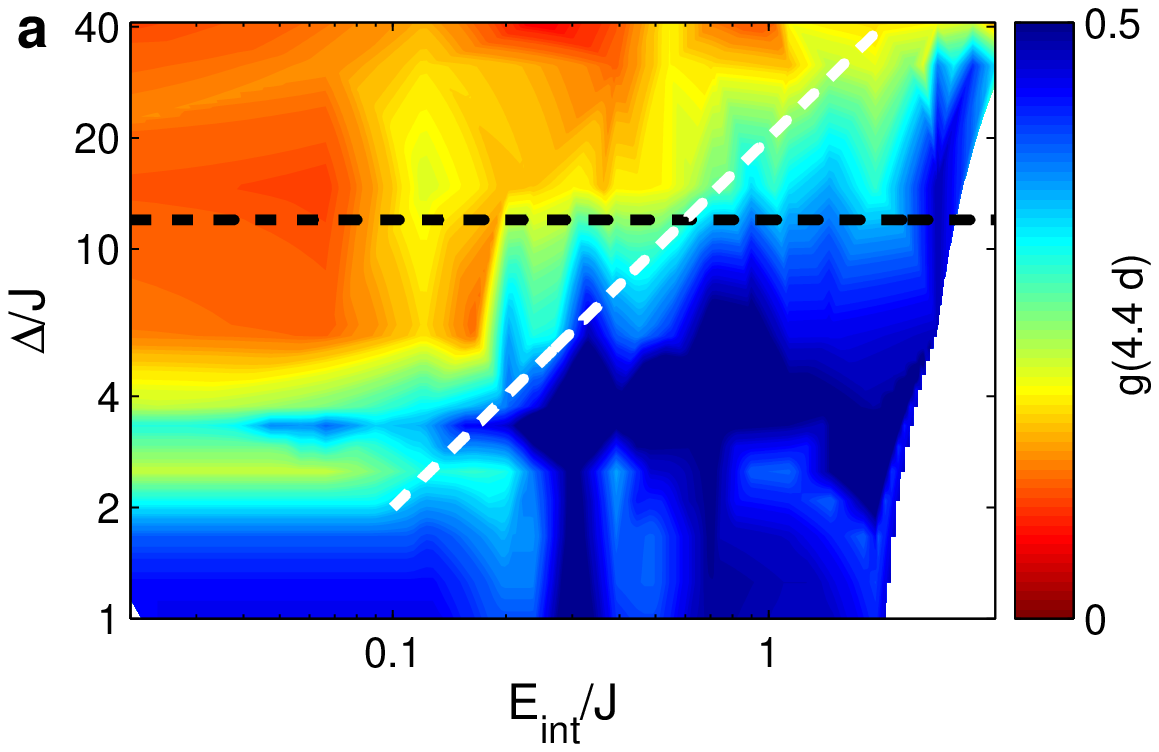}
\includegraphics[width = 0.33\textwidth]{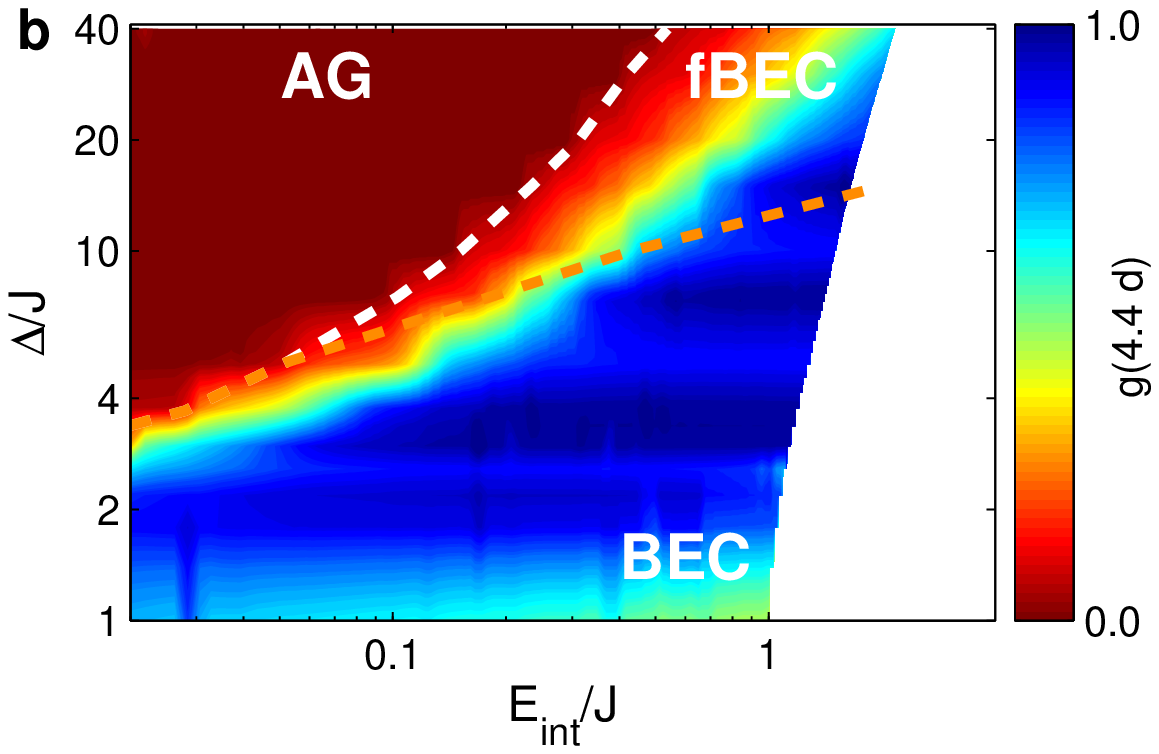}
\includegraphics[width = 0.33\textwidth]{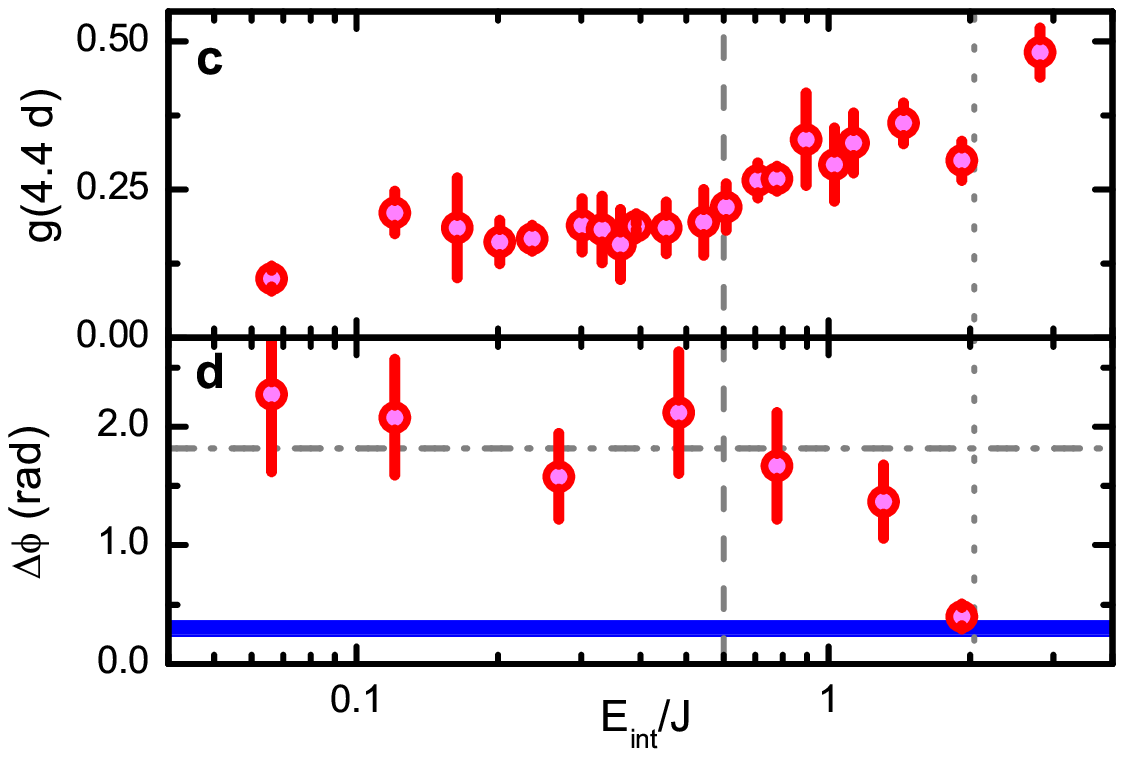}
\end{center}
\caption{\capbf{Probing the phase coherence of the system.} \capbf{a \& b}, Correlation of neighbouring localized states. \capbf{a}, Experiment (white line as in Fig.~3), \capbf{b}, Theory of the ground state. The white and orange dashed lines show the phase boundaries between an Anderson glass (AG), a fragmented BEC (fBEC), and a macroscopic Bose-Einstein condensate (BEC). \capbf{c}, Correlation of neighbouring states for $\Delta/J = 12$, corresponding to black dashed line in \capbf{a}. The error bar is given by the standard error of the mean. \capbf{d}, Standard deviation of the phase measured by repeating the experiment up to 25 times for a given set of parameters for $\Delta/J = 12$. The error is estimated as $\Delta\phi/\sqrt{N}$, where $N$ is the number of images from which the phase was extracted. The blue bar shows the phase fluctuations measured for an extended system below the localization threshold. The grey dash-dotted line gives the standard deviation for a purely random distribution. In \capbf{c} \& \capbf{d}, the grey dashed (dotted) lines give $E_{int} = 0.05\Delta$ ($0.17\Delta$).}
\label{fig:phase}
\end{figure*}

The system is characterized in detail by analyzing its momentum distribution, which is recovered by taking an image of the condensate after a long ballistic expansion without interactions (see Methods). From the momentum distribution and derived Fourier transforms, of which we show examples in Fig.~2, we extract the local shape of the wavefunction, spatial correlations, and phase coherence properties for different values of $\Delta/J$ and $E_{int}/J$. The system can be approximately described as the superposition of states with the same envelope separated by $4.4d$. First, the mean extension of individual states can be quantified by measuring the root-mean-squared width of the momentum distribution (Fig.~2a-b). A smaller (larger) width indicates a more extended (localized) state. 
Next, the mean local shape of the wavefunction on a length scale of $4.4d$ is extracted from the Fourier transform of the square root of the momentum distribution. From a fit to a generalized exponential function, the localization exponent $\alpha$ is recovered (see also Methods), as shown in Fig.~2c-d.
The measured momentum width and exponent are shown in Fig.~3. We find that for very small $E_{int}$, the states are exponentially localized, since $\alpha\approx 1$, and the momentum width is large, consistent with the Anderson glass regime. Increasing $E_{int}$, the width decreases while the exponent increases up to $\alpha\approx 2$. Repulsive interactions therefore delocalize the system as expected, or alternatively, the localization transition is shifted to higher values of the disorder strength $\Delta/J$ when interactions are introduced into the system. 
The position of the delocalization crossover is in good agreement with the expectations of a simple screening argument\cite{Lee90}: The increasing interaction energy serves to smooth over the disordering potential in the occupied sites, providing a flatter energetic landscape on which more extended states can form. 
The centre of the crossover is therefore expected to occur when $E_{int}$ is comparable to the standard deviation of energies in the lowest miniband of the non-interacting spectrum, $0.05\Delta$ (white line in Fig.~3, see also Supplementary Information).

The correlation properties of neighbouring states can be extracted from a Fourier transform of the momentum distribution itself, which gives the spatially averaged first order correlation function $g(x)$ (see Methods). In Fig.~4a-b, $g(x)$ at $x=4.4d$ is shown for both the experiment and a ground-state theory that we have developed, with generally good agreement. In the localized regime, the correlation is exactly zero in the theory, since no neighbouring states are occupied. In contrast, the correlation is finite in the experiment due to the occupation of neighbouring localized states arising from the non-adiabatic loading, but is small since the states are independent. As $E_{int}$ is increased, the correlation features a crossover towards larger values, signalling that coherence is progressively established locally over distances of at least $4.4d$. The shape of the crossover in the experiment is again in qualitative agreement with the expectation of the screening argument above (see white line in Fig.~4a).

Finally, information about the phase coherence of neighbouring states can be obtained by measuring the phase $\phi$ of the interference pattern in the momentum distribution for repeated runs of the experiment with the same parameters (see Methods for details). If the states are not phase locked, $\phi$ changes almost randomly at each repetition of the experimental sequence. 
In Fig.~4d we show the standard deviation of $\phi$, estimated from a large number of repetitions of the experiment, for fixed $\Delta/J=12$. We see a slight decrease of the phase fluctuations with increasing $E_{int}$, that nevertheless remain relatively large in the crossover region where the correlation increases (Fig.~4c). The fluctuations finally drop to the background value only when $E_{int}$ is comparable to the full width of the lowest miniband of the non-interacting spectrum, $0.17\Delta$. These observations confirm that in the localized regime the states are totally independent, which together with the localization properties (Fig.~3) indicates that the system can indeed be described as an Anderson glass\cite{Scalettar91,Damski03}. The system crosses a large region of only partial coherence while becoming progressively less localized as $E_{int}$ is increased. This is consistent with the formation of locally coherent fragments expected for a fBEC. An analogous fragmentation behaviour was reported in ref.~\citen{Chen08}. Finally, the features of a single extended, coherent state are seen, i.e.\ a BEC. 

In the mean-field theory, boundaries between the different phases expected for the system can be defined (see Methods). In particular, the transition from the Anderson glass phase to a fragmented BEC (white line in Fig.~4b) occurs when $g(4.4d)$ starts to increase. Similarly, the orange line in Fig.~4b shows where the fragments are locked together in phase to form a single macroscopic condensate for very large interactions. The generally good agreement between the experimental observables and theory indicates that our system is well described by the mean-field theory for most of the parameter space explored experimentally.

In conclusion, we have provided the first experimental characterization of the localization, correlation and coherence properties of the various regimes due to the competition of disorder and weak repulsive interactions in a bosonic system. Other aspects of the delocalization crossover worth further study are, e.g.\ the detailed properties of the ground state of the AG regime, which was not possible to study in the present set-up, and the presence of a superfluid-insulator transition at the BEC-fBEC boundary analogous to the one observed in superconductors\cite{Dubi07}. Regarding the latter, in transport experiments analogous to the ones described in ref.~\citen{Roati08} we have been able to verify that the AG and fBEC regimes are not inconsistent with being insulating, as is the case in the regime of vanishing $E_{int}$ (see Supplementary Information). Finally, it would be appealing to employ the present system and the correlation analysis introduced here to explore the regime of strong correlations, $E_{int} \gg J$, which could be reached by using a quasi-1D system with strong radial confinement. There, another elusive insulating phase due to the \emph{cooperation} of disorder and interactions, the so called Bose-glass phase, is expected to appear, although there is debate on the exact shape of the phase diagram\cite{Fisher89,Scalettar91,Damski03,Fallani07}.

\begin{methods}
\subsection{Condensate with tunable interactions.}
A ${}^{39}$K condensate of about $N=20,000$ atoms with an $s$-wave scattering length of $250a_0$, where $a_0=52.9$~pm is the Bohr radius, is prepared in a harmonic optical trap. The condensate is loaded into the quasi-periodic potential while the optical trap is decompressed in about 250 ms to reduce the harmonic confinement, and a gravity-compensating magnetic field gradient is added. At the same time, the scattering length $a$ is changed by means of a broad Feshbach resonance to values ranging from $a\leq 0.1 a_0$ to about $a =300 a_0$ (ref.~\citen{Roati07}).

\subsection{Quasi-periodic potential.}
The quasi-periodic potential is created by two vertically oriented laser beams in standing-wave configuration. The primary lattice is generated by a Nd:YAG laser with a wavelength of $\lambda_1 = 1064.4$~nm and has a strength of $s_1 = V_1/E_{\mathrm{R},1} = 10.5$ (corresponding to $J/h = 79$~Hz), as measured in units of the recoil energy $E_{\mathrm{R},1} = h^2/(2M\lambda_1^2)$. The secondary lattice is generated by a Ti:Sapphire laser of wavelength $\lambda_2 = 866.6$~nm, the strength being adjustable up to $s_2 = V_2/E_{\mathrm{R},2} = 1.7$. Both beams are focussed onto the condensate with a beamwaist of about 150~$\mu$m. The lattice lasers give a harmonic confinement of $\omega_\perp = 2\pi \times 50$~Hz in the radial direction. In the vertical (axial) direction, a weak confinement of 5~Hz is given by a weak optical trap as well as by a curvature from the gravity-compensating magnetic field.

\subsection{Energy scales.}
In the tight-binding limit, the hopping energy $J$ and disorder strength $\Delta$ can be estimated as $J = 1.43 s_1^{0.98} \exp\left\{-2.07\sqrt{s_1}\right\} E_{\mathrm{R},1}$ and $\Delta = 0.5 s_2 \beta^2 \left( 1.0264 \exp \left\{-2.3624/s_1^{0.59} \right\} \right) E_{\mathrm{R},1}$ (ref.~\citen{Modugno09}). The experimental uncertainty on $\Delta/J$  is around 15\%. We estimate that around 30 lattice sites, corresponding to about 7 localized states, are populated during the loading of the lattice. We then define a mean interaction energy per particle $E_{int} = gN/7 \int |\varphi(\mathbf{r})|^4 \, d^3\mathbf{r}$, where $g = 4\pi\hbar^2a/m$ and $\varphi(\mathbf{r})$ is a Gaussian approximation to the on-site Wannier function. We include coupling into the radial directions of our system, with the consequence that the interaction energy is non-linear in the scattering length. Though this definition of the energy is strictly valid only in the localized regime, comparison with a numerical simulation of our experimental procedure has shown that it is a good approximation for all values of the scattering length up to an error of 30\%. Note that the potential energy from the residual harmonic confinement is approximately $3\times10^{-3}J$ over a distance $4.4d$.

\subsection{Momentum distribution analysis.}
The images of the momentum distribution are taken by absorption imaging with a CCD camera after 36.5~ms ballistic expansion. At the time of release, the scattering length is set to below $1a_0$ in less than one ms and kept there until the Feshbach magnetic field is switched off 10~ms before taking the image -- at this point, the system has expanded a sufficient amount to minimize the effect of interactions. For such a free expansion, the image is approximately the in-trap momentum distribution $\rho(k) = \langle \hat\Psi^\dagger(k) \hat\Psi(k) \rangle$ (ref.~\citen{Bloch08}). The acquired images are integrated along the radial direction to obtain a profile. In momentum space, the width of the central peak is calculated by taking  the root mean square width within the first Brillouin zone.
Due to the quasi-periodic lattice potential, for a sufficiently homogeneous system the in-trap wavefunction can be decomposed into copies of a single state with real and non-negative envelope $\xi(x) \sim \exp(-|x/L|^\alpha)$, spaced by $4.4d$. Therefore in momentum space, $\sqrt{\rho(k)} = \xi(k) \mathcal{S}(k)$, where $\mathcal{S}(k)$ is an interference term, and $\xi(x)$ can be extracted from a Fourier transform of $\sqrt{\rho(k)}$ (see also Supplementary Information). We fit to the sum of two generalized exponential functions $A\exp\left(-|(x-x_c)/L|^\alpha\right) \cdot [1+B\cos(k_1 (x-x_c) + \delta)]$, where $x_c$ denotes the centers of each of these functions, spaced by $4.4d$. From this fit, the exponent $\alpha$ is recovered.
In addition, from the Wiener-Khinchin theorem, the momentum distribution can be expressed in terms of the first order correlation function $G(x',x+x') = \langle \hat \Psi^\dagger(x') \hat \Psi(x+x') \rangle$ as $\rho(k) \propto \mathfrak{F}^{-1} \int G(x',x+x') \, dx'$. By taking the Fourier transform of the momentum distribution itself, we can therefore recover the spatially averaged correlation function $g(x) = \int G(x',x+x') \, dx'$. With the same fitting function as above, we evaluate the spatially averaged correlation between two states 4.4 lattice sites apart, $A_2/A_1$. Experimentally, the correlation function saturates at a value around 0.5 due to the finite momentum resolution. 
The fluctuations in phase between neighbouring states are seen as a fluctuation of the phase $\phi$ of the interference pattern of the momentum distribution, which is directly extracted from a fit (see also Supplementary Information).
The 2D graphs in Figs~3 and 4 were generated by linearly interpolating a total of 130 averaged datapoints at 9 different values of disorder, changing the interactions. Typical experimental scatter and statistical errors are seen in Fig.~4c.

\subsection{Theory of the ground state.}
The theoretical calculations presented in the paper rely on a meanfield approach similar to the one of ref.~\citen{Salasnich02}. This is an effective one-dimensional model which partially includes also the radial to axial coupling, and is known
to provide an accurate description in the two limiting cases of Anderson localization and BEC. The boundaries between the different regimes shown in Fig.~4b are obtained by analyzing the correlation function $g(x)$ and the density distribution. In the theory we define the Anderson glass phase as the one in which the correlation $g(4.4d)$ is zero. 
To enter the fragmented BEC (fBEC) phase, we require $g(4.4d)>0$, which implies that coherent fragments composed of adjacent localized states can start to form. 
For increasing $E_{int}$ the extension of the fragments increases, until most of the system remains in a single component, which corresponds to a macroscopic BEC. To define the boundary between the fBEC and the BEC regimes, we first identify as fragments the parts of the system separated by low-density regions for which an applied relative phase twist does not affect the energy of the system. When one single macroscopic fragment forms, we assume the system to be in the BEC regime. A more detailed description of the theoretical methods can be found in the Supplementary Information.

\end{methods}

\begin{addendum}
 \item[Acknowledgements] We thank B. Altshuler and M. Larcher for discussions, S. M\"uller for experimental contributions and all the colleagues of the Quantum Gases group at LENS. This work has been supported by the EU (MEIF-CT-2004-009939), by the ESA (SAI 20578/07/NL/UJ), by the ERC through the Starting Grant QUPOL and by the ESF and CNR-INFM through the EuroQUASAR program.
 \item[Additional Information] Correspondence and requests for materials should be addressed to B.D. (email: deissler@lens.unifi.it).
\end{addendum}


\clearpage

\renewcommand{\thefigure}{S\arabic{figure}}
\renewcommand{\theequation}{S\arabic{equation}}
\setcounter{figure}{0}
\setcounter{equation}{0}

\section*{\LARGE Supplementary Information}

\subsection{Energy spectrum}
\begin{figure}[b]
\begin{center}
\includegraphics[width=0.45\textwidth]{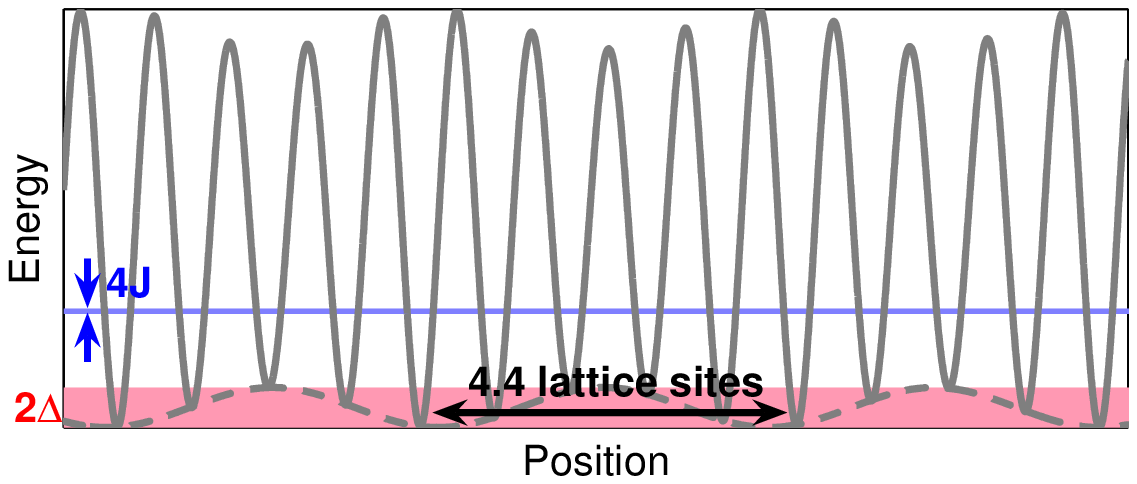}
\caption{\capbf{Quasi-periodic potential.} The quasi-periodic potential realized in the experiment for lattice incommensurability $\beta = 1.228\ldots$ and $\Delta/J = 15$. The red stripe shows the amplitude $2\Delta$ of the perturbation by the secondary lattice. The blue stripe denotes the position of the first Bloch band of the primary lattice, with width $4J$. The quasi-periodic lattice is characterized by potential wells approximately every $1/(\beta-1) \approx 4.4$~lattice sites, which arise from the beating of the two lattices (dashed grey line).}
\label{fig:bichromaticS}
\end{center}
\end{figure}

The potential of our system can be written in general as
\begin{equation}\label{eq:potential}
    V(x)=s_{1}E_{R1} \sin^{2}(k_{1}x)+s_{2}E_{R2} \sin^{2}(k_{2}x+\phi) + V_{ext}(x,\mathbf{r}_\perp),
\end{equation}
where $E_{\mathrm{R},i}=\hbar^{2}k^{2}_{i}/(2M)=h^{2}/(2M\lambda^{2}_{i})$ is the recoil energy for the lattice with wavelength $\lambda_{i}=2\pi/k_{i}$, and $s_{i}=V_{i}/E_{\mathrm{R},i}$  is the height of the lattice $i$ in units of $E_{\mathrm{R},i}$. Each of the two lattices is $\lambda_{i}/2$ periodic. Any external confining potential is given by $V_{ext}(x,\mathbf{r}_\perp)$. The lattice spacing of such a potential is to good approximation $d = \lambda_1/2$. If the ratio $\beta = k_2/k_1$ is an irrational number, eq.~\ref{eq:potential} describes a quasi-periodic potential. In our case, $\lambda_1 = 1064.4$~nm and $\lambda_2 = 866.6$~nm, giving $\beta \approx 1.228$.

The essential features of such a potential are visible in Fig.~\ref{fig:bichromaticS}. The potential energy minima of the primary lattice are modulated by the second one, giving rise to characteristic wells separated on average by $1/(\beta-1) \approx 4.4$ lattice sites. The energy scales that characterize the corresponding Hamiltonian to such a potential are the tunnelling energy of the primary lattice\cite{SGerbier05}
\begin{equation}
J = 1.43 s_1^{0.98} \exp\left\{-2.07\sqrt{s_1}\right\} E_{\mathrm{R},1},
\end{equation}
and the disorder energy\cite{SModugno09}
\begin{equation}
\Delta = 0.5 s_2 \beta^2 \left( 1.0264 \exp \left\{-2.3624/s_1^{0.59} \right\} \right) E_{\mathrm{R},1}.
\end{equation}

\begin{figure}[!t]
\begin{center}
\includegraphics[width=0.33\textwidth]{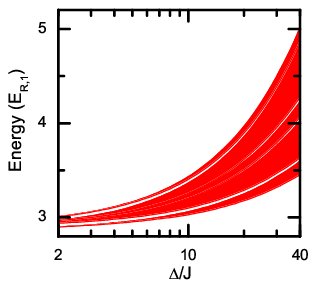}
\caption{\capbf{Energy spectrum of the quasi-periodic potential.} Energies of eigenstates of the non-interacting system as a function of $\Delta/J$.}
\label{fig:Espectrum}
\end{center}
\end{figure}

Neglecting the external confining potential, the spectrum of such a quasi-periodic potential can easily be calculated and is shown in Fig.~\ref{fig:Espectrum} for various values of the disorder strength $\Delta/J$. A striking feature is the appearance of minigaps in the spectrum, the lowest of which has approximately the same width for all values of $\Delta/J$. A minigap appears when the potential has two neighbouring lattice sites with almost the same minimum potential energy. Locally, the potential then looks like a double well, for which the two lowest-lying eigenstates have an energy splitting of $2J$. In fact, the width of the lowest minigap is approximately $2J$ throughout the range of $\Delta/J$ shown. The lowest ``miniband'' of energies corresponds to the lowest energy eigenstates localized in the potential wells $4.4d$ apart. Since in the experiment, only the states in the first ``miniband'' are populated, we restrict our analysis to these energies and find that their standard deviation is approximately $0.05\Delta$, while the extension of this band is approximately $0.17\Delta$. The effect of a confining potential on the spectrum has been analysed previously in ref.~\citen{SModugno09}.

\subsection{Momentum distribution analysis}
Fourier transform techniques are used to extract information both about the local shape of the wavefunction, and about the coherence properties of neighbouring states. After a long free expansion without interactions, the image of the atoms that is acquired is approximately the in-trap momentum distribution $\rho(k) = \langle \hat\Psi^\dagger(k) \hat\Psi(k) \rangle$, where $\hat\Psi(k)$ is the Fourier transform of the bosonic field operator $\hat\Psi(x)$. In order to recover information about the in-trap wavefunction, we can therefore use an inverse Fourier transform.

\begin{figure}
\begin{center}
\includegraphics[width=0.33\textwidth]{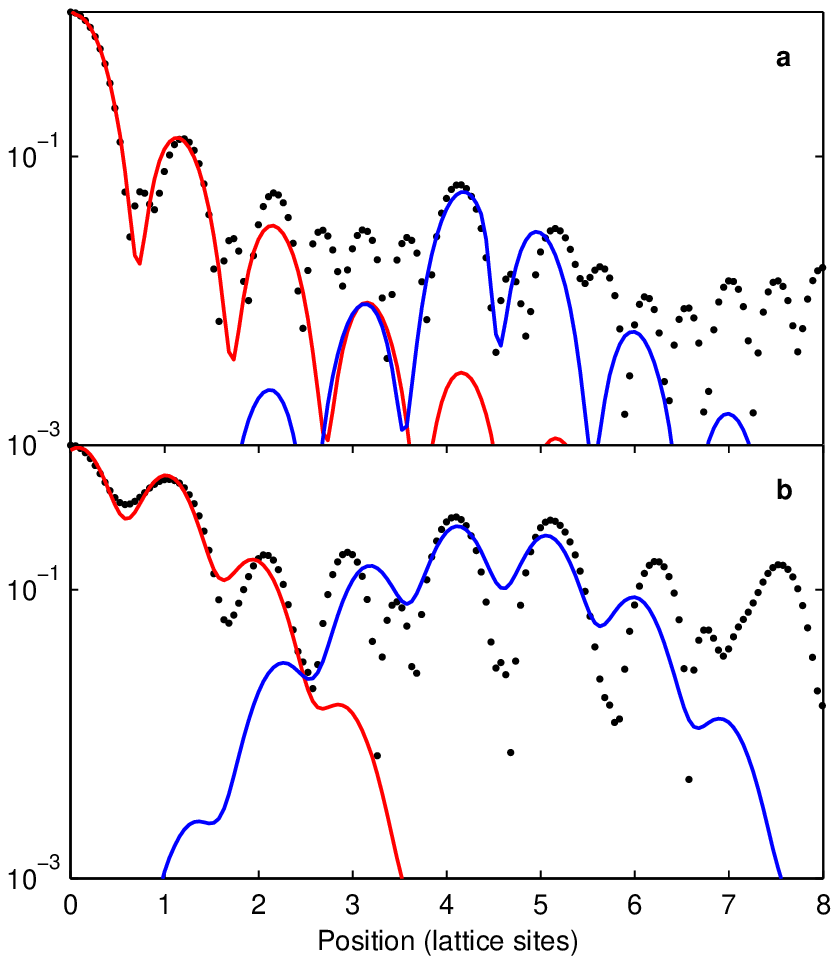}
\caption{\capbf{Examples of Fourier analysis.} Typical Fourier transforms (dots) of the square root of the momentum distribution giving the average local shape of the wavefunction, \capbf{a}, in the localized regime, \capbf{b}, for a coherent, extended state. The red and blue lines give the fits to the data for a generalized exponential function centered at the origin (red line) and at $4.4d$ (blue line).}
\label{fig:FFTexamples}
\end{center}
\end{figure}

Due to the quasi-periodic nature of the employed lattice potential, we expect that for a sufficiently homogeneous system, the in-trap wavefunction can be decomposed into copies of the same state with real and non-negative envelope $\xi(x)$, spaced by $D = 4.4d$. The overall wavefunction can therefore be approximated as
\begin{equation}
 \psi(x) = \sum_j  a_j \xi(x-jD) e^{-i\phi_j}, 
\end{equation}
where $\phi_j$ is the local phase, and a useful example of $\xi(x)$ is a generalized exponential function $\exp(-|x/L|^\alpha)$. In momentum space, the magnitude of the overall wavefunction can then be written as $\sqrt{\rho(k)} = |\xi(k)| \mathcal{S}(k)$, where
\begin{equation}
 \mathcal{S}(k) = \left| \sum_j a_j e^{-i(jkD + \phi_j)} \right| 
\end{equation}
is an interference term. For many envelope functions $\xi(x)$, such as the generalized exponentials with $0 < \alpha \leq 2$, the Fourier transform $\xi(k)$ itself is real and non-negative\cite{SGiraud06}, so that the inverse Fourier transform of $\sqrt{\rho(k)}$ can be written as $\xi(x) \circ \mathcal{S}(x)$. This is simply the convolution of the envelope of a single state $\xi(x)$ with the Fourier transform of the interference term, $\mathcal{S}(x)$, which can be approximately described as a series of sharp peaks (approaching $\delta$-distributions) spaced by $D$, with a decreasing amplitude and phases that depend on the local phases $\phi_j$ and amplitudes $a_j$. 

\begin{figure}[t]
\begin{center}
\includegraphics[width=0.45\textwidth]{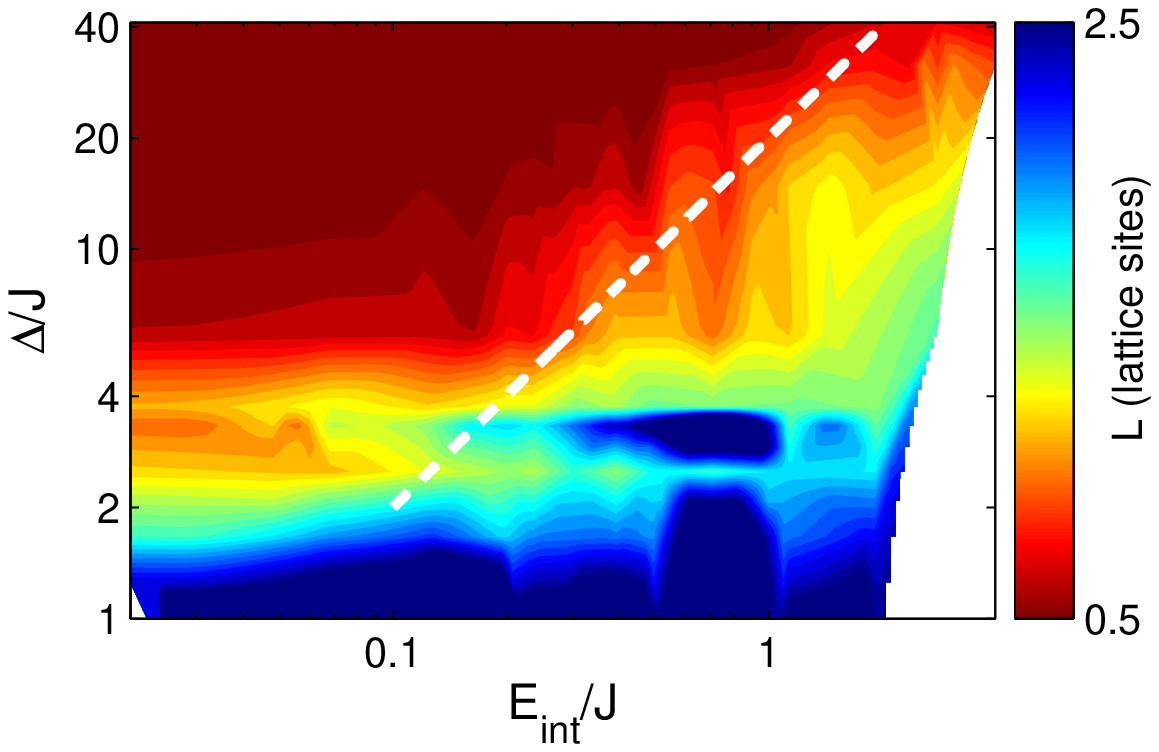}
\caption{\capbf{Local size of states.} The local size $L$, as extracted from fits to the local shape of the wavefunction, gives the average extension of the individual states. This quantity is strictly valid only in the localized regime. The white line gives the expected position of the delocalization crossover ($E_{int} = 0.05\Delta$).}
\label{fig:locallength}
\end{center}
\end{figure}

The inverse Fourier transform of the square root of the momentum distribution $\rho(k)$ therefore gives the average local shape of the (wave)function $\xi(x)$. Due to our finite resolution in momentum space (about $k_1/20$), we are only able to resolve easily two neighbouring states. The averaged wavefunction is analysed by fitting to the sum of two generalized exponential functions modulated by the primary lattice
\begin{equation}
\sum_{j=1,2} A_j \exp\left(-|(x-x_j)/L|^\alpha\right) \cdot \left[ 1+B\cos(k_1 (x-x_j) + \delta_j) \right],
\label{eq:expfit}
\end{equation}
where $x_1 =0$ and $x_2 = 4.4d$ (see Fig.~\ref{fig:FFTexamples} for examples). From such a fit, the exponent $\alpha$ can be extracted. The local size $L$ is shown in Fig.~\ref{fig:locallength}, which shows a complementary behaviour to the momentum width (Fig.~3a in the main text).

On the other hand, the inverse Fourier transform of the momentum distribution itself can be employed to find the correlation properties of neighbouring states. Using the Wiener-Khinchin theorem, the momentum distribution $\rho(k)$ can be related to the first order correlation function $G(x',x+x') = \langle \hat \Psi^\dagger(x') \hat \Psi(x+x') \rangle$, as $\rho(k) \propto \mathfrak{F}^{-1} \int G(x',x+x') \, dx'$. By taking the Fourier transform of the momentum distribution, we can therefore recover the spatially averaged correlation function $g(x) = \int G(x',x+x') \, dx'$. We fit with the same generalized exponential of Eq.~\ref{eq:expfit} and recover the spatially averaged correlation between two states 4.4 lattice sites apart as $A_2/A_1$. Also here, the finite momentum resolution limits our analysis to two neighbouring sites, and it follows that the correlation function saturates at a value around 0.5.

The effect of a fluctuating phase between neighbouring states is seen as a shift of the phase $\phi$ of the interference in the momentum distribution. We extract this phase by fitting the momentum distribution directly with a fitting function
\begin{equation}
\sum_{\tilde{k} = k_C,k_C \pm 2k_1} A_{\tilde{k}} \exp\left(-(k-\tilde{k})^2/2w^2\right) \cdot \left[ 1+B\cos\left(D(k-\tilde{k}) + \phi \right) \right],
\end{equation}
where $k_C$ is the center of the distribution, determined by fitting the average of all images of a given dataset.

\subsection{Adiabaticity}

\begin{figure}
\begin{center}
\includegraphics[width=0.45\textwidth]{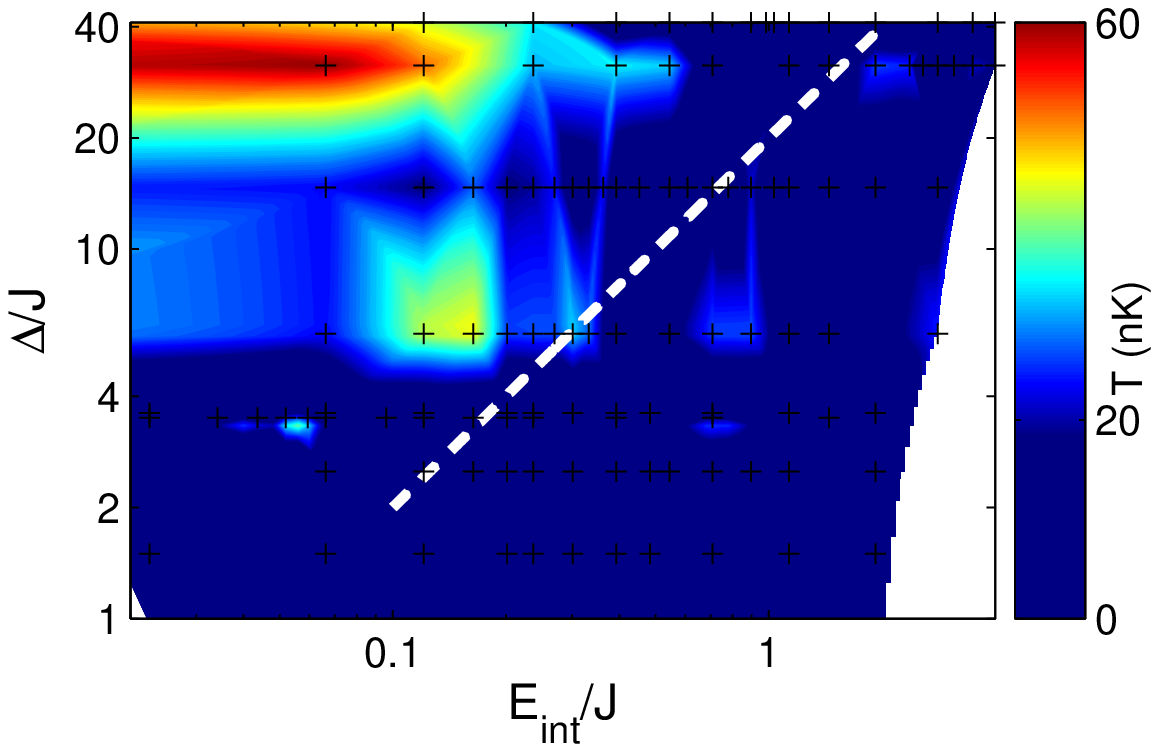}
\caption{\capbf{Adiabaticity of preparation.} The radial temperature is estimated from radial profiles extracted from the images. The white line gives the expected position of the delocalization crossover ($E_{int} = 0.05\Delta$). The black crosses show the positions in the $\Delta/J$-$E_{int}/J$ plane of the datapoints from which all 2D graphs were interpolated.}
\label{fig:radial}
\end{center}
\end{figure}

As discussed in the paper, in the experiment the system is prepared by first loading a strongly interacting condensate ($a=250 a_0$) into the quasi-periodic potential, in presence of a tight axial harmonic confinement (60 Hz). The scattering length is then reduced to its final value with an exponential ramp lasting 250 ms ($\tau \approx 25$~ms), while the harmonic confinement is linearly reduced to a much lower value (5 Hz). We check the degree of adiabaticity of this procedure by monitoring the evolution of the radial degrees of freedom. 
From the radial profiles extracted from the absorption images, we can measure the radial temperature of the system. We find that there is always either a quasi-pure condensate or condensed and thermal components. In the latter case, the temperature can be directly measured from the thermal component, while in the former case, only an upper bound for the temperature (about 20~nK) can be given. The condensation temperature $T_C$ varies across the parameter range explored, but we estimate\cite{SBurger02} that it is on the order of 60--100~nK. 

We interpret the radial excitation as an interaction-mediated transfer of excitation energy from the axial to the radial degrees of freedom. What presumably happens in the system is the following: As the tunnelling time between the states separated by $4.4d$ becomes longer than the experimental timescale for decreasing the interaction strength, the population of these states cannot move adiabatically to the absolute ground state of the system. The states are therefore left with an excess population, hence an excess interaction energy that spatially broadens them by populating even more excited states in the neighbouring sites of the primary lattice. These excited states have an energy on the order of the mean separation of energies of neighbouring lattice sites $0.8\Delta$, and can decay to the lowest state in the potential well on a timescale of $\Delta^{-1}$ by transferring their excess energy to the radial directions.

The observation that this process takes place only near the onset of the Anderson glass regime is in agreement with the expectation of suppressed tunnelling between the states $4.4d$ apart in the exponentially localized regime. A radial excitation therefore shows that there are axial excitations, e.g. due to a lack of adiabaticity during the loading procedure. On the other hand, a lack of radial excitations is not strictly a proof that there are no axial excitations, since the radial heating is caused by \emph{local} axial excitations, i.e. arising from neighbouring lattice sites, while there could be axial excitations over longer distances that are not able to release energy by relaxing into the true ground state.

\subsection{Ground state phase diagram}
\begin{figure}[t]
\begin{center}
\includegraphics[width=0.5\textwidth]{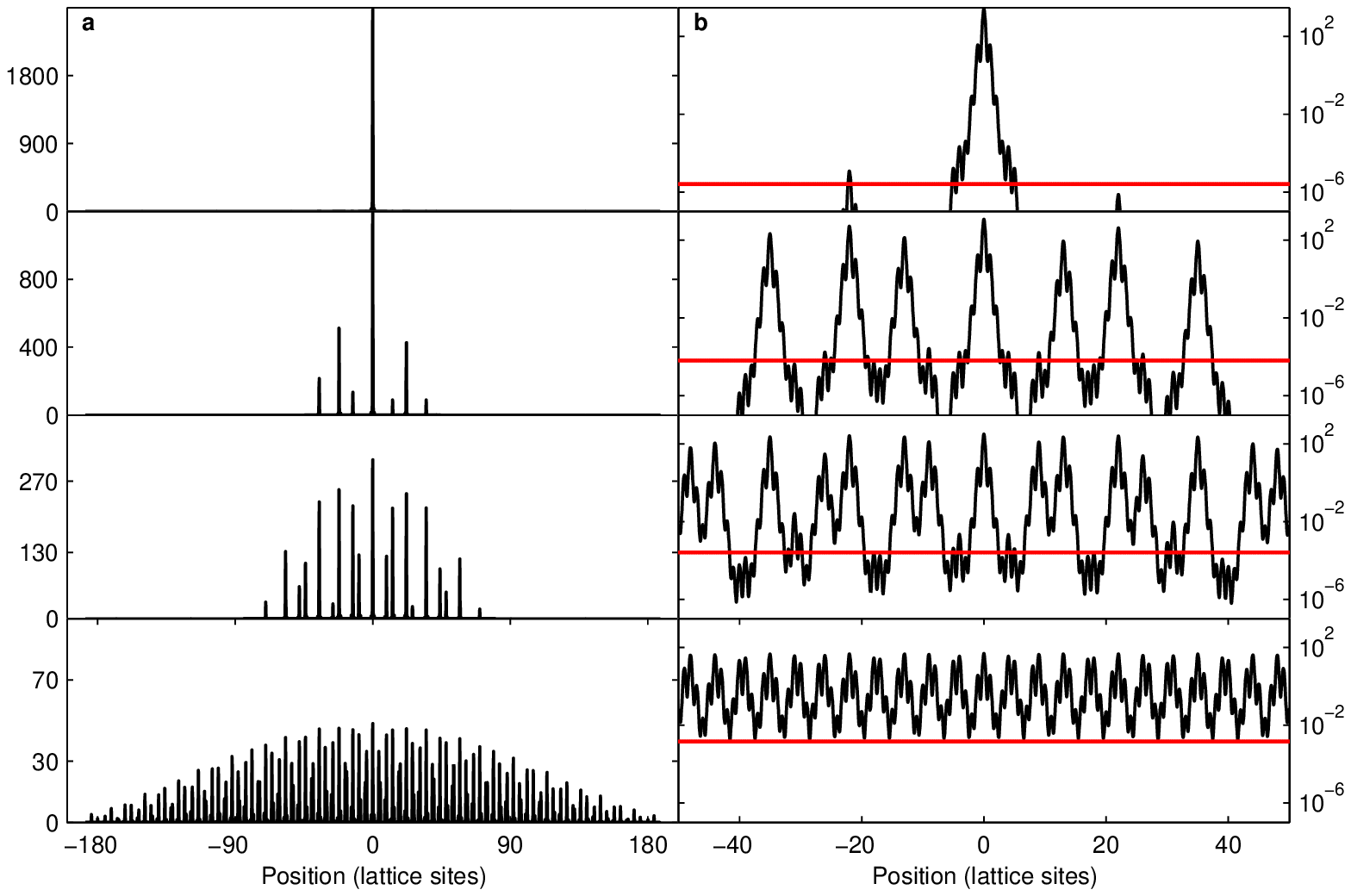}
\caption{\capbf{Theoretical density distribution.} Density distribution calculated from the ground state theory for $\Delta/J = 10$, \capbf{a}, on a linear scale, and \capbf{b}, on a logarithmic scale. The interaction energy increases from top to bottom ($E_{int}/J = 0.003, 0.066, 0.26, 1.5$). The top two panels are for $E_{int}$ such that the system is an Anderson glass, the third panel shows a fragmented BEC, and the lowest panel depicts a BEC. The red lines show $E_{int}/N$, which is used to evaluate the different phases.}
\label{fig:theory}
\end{center}
\end{figure}

For the calculation of the theoretical profiles and ground state phase diagram presented in this paper we have used the following meanfield approach. In the limiting cases of a single Anderson localized state (AL) or of a coherent condensate (BEC), the system can be described by a single wavefunction $\psi(x,\mathbf{r}_\perp)$ that is a solution of the Gross-Pitaevskii equation in the BEC regime, or of the Schr\"odinger equation in the AL regime. In the latter case the problem is separable, and the wavefunction can be factorized as $\psi(x,\mathbf{r}_\perp)=\varphi(x)\phi(\mathbf{r}_\perp)$.
In the intermediate regime of a glassy phase made of several independent fragments, the atomic distribution that minimizes the energy can be accounted for by considering a wavefunction of the form $\psi(x,\mathbf{r}_\perp)\approx\sum_i\varphi_i(x)\phi_i(\mathbf{r}_\perp)$ (similarly to the approach used by Lugan \textit{et al.}\cite{SLugan07}). An approximate way to describe the overall behaviour of the system in the crossover between different regimes is to consider an \textit{effective wavefunction} $\psi(x,\mathbf{r}_\perp)=\varphi(x)\phi(\mathbf{r}_\perp, \sigma(x))$, where the radial component is a Gaussian with an $x$-dependent width, and minimizing the corresponding energy functional
\begin{align}\label{eq:Efunc}
E\left[\varphi,\sigma\right] &=\int dx \, \varphi^*(x)
\left[
-\frac{\hbar^2}{2m}\nabla^2_x
+V(x)
+\frac{\hbar^2}{2m}\frac{1}{\sigma^2}
+\frac{m\omega_\perp^2}{2}\sigma^2\right]\varphi(x) \nonumber\\
&+ \int dx \, \frac{1}{2}\frac{gN}{2\pi\sigma^2}|\varphi|^4 ,
\end{align}
with $V(x)$ given by eq.~\ref{eq:potential},
that corresponds to solving the non-polynomial Schr\"odinger equation of ref.~\citen{SSalasnich02}.
This approach has the advantage of capturing the modification of the axial density distribution in the crossover from the localized to the extended regime, and we use it to evaluate the spatial and momentum distributions of the ground state of the system (Fig.~\ref{fig:theory}). In particular, from the calculated momentum distributions we extract the correlation function $g(4.4d)$  shown in Fig.~4b as we do for the experimental data. 

The boundaries between different regimes in Fig.~4b have been derived by studying both the $g(4.4d)$ and the spatial distributions. For example, the system is expected to pass from the Anderson glass (AG) to the fragmented BEC (fBEC) when neighbouring localized states start to be macroscopically occupied and have a relative phase coherence. On the one hand, a non-zero value of $g(4.4d)$ indicates the occupation of neighbouring states. On the other hand, we can check for the coherence properties of such states by studying the effect of introducing a phase twist over an extension $d$ between them.  If the two states are independent, the phase twist will not modify the energy of the system, while the contrary happens if they are coherent.  In practice, we evaluate the energy cost per particle, 
\begin{equation}
\delta E_{k}\approx\frac{\hbar^2}{2 m}\frac{(2\pi)^2}{d}|\varphi(x_k)|^2,
\end{equation}
for introducing a $2\pi$ phase twist at the lattice site $k$ where the two states connect. If $\delta E_{k}$ is less than the change in the energy per particle associated with the removal of one atom from the system, $E_{int}/N$, where $E_{int}$ is the last term in eq.~\ref{eq:Efunc}, then the two states are considered independent. We define the boundary between the AG and fBEC regimes where $g(4.4d)$=0.01. We have verified that above this threshold the neighbouring states are also phase locked, i.e. they constitute a coherent fragment. For increasing interaction energy the size of these coherent fragments increases, while their number decreases. Due to the harmonic confinement, the system tends to form a large central core surrounded by fragments in the low-density tails. We assume that the system has entered the BEC regime when there is a single macroscopic fragment at the center of the trap, and the population of each of the outer fragments is less than $1\%$ of that of the central one.

\subsection{Transport experiments}

In the experiment, in addition to studying the momentum distribution, we have investigated the transport properties of the various regimes of the system, using the same technique we already employed in ref.~\citen{SRoati08}. The technique consists of suddenly releasing the axial harmonic confinement, while keeping both the quasi-periodic lattice potential and the radial confinement, and then observing the subsequent diffusion (or lack thereof) of the atomic cloud in the axial direction. In the previous experiment we observed that for vanishing $E_{int}$ the diffusion becomes strongly suppressed for $\Delta/J>2$. We have now observed that it continues to be suppressed for $\Delta/J \gtrsim 2$ also for the values of $E_{int}$ explored in the present work, irrespective of whether the system is in the AG, fBEC or BEC regimes. We can interpret the absence of diffusion in the AG (fBEC) regime as a result of the insulating nature of the system, which occupies (partially) localized states. In contrast, in the extended BEC regime, a slow subdiffusive expansion is expected\cite{SLarcher09}, or the expansion might be completely suppressed for some values of the parameters by self-trapping\cite{SAnker05}. The slow expansion however would probably not be detectable on the one second timescale of the experiments we have performed so far. The absence of a different diffusion behaviour of the system in the AG, fBEC and BEC regimes unfortunately does not allow us to identify the boundary between the fBEC and BEC regimes using this technique, nor to unambiguously prove the insulating nature of the AG and fBEC regimes. Other methods therefore need to be developed for such a purpose.

\section*{Supplementary References}

\end{document}